\documentstyle[epsf]{article}
\newcommand{\bee}{\begin{equation}}
\newcommand{\ee}{\end{equation}}
\newcommand{\beea}{\begin{eqnarray}}
\newcommand{\eea}{\end{eqnarray}}

\begin{document}
\thispagestyle{empty}
\parskip=12pt
\raggedbottom

\def\mytoday#1{{ } \ifcase\month \or
 January\or February\or March\or April\or May\or June\or
 July\or August\or September\or October\or November\or December\fi
 \space \number\year}
\noindent
\hspace*{9cm} COLO-HEP-393\\
\vspace*{1cm}
\begin{center}
{\LARGE  Revealing  Topological Structure in the $SU(2)$ Vacuum}

\vspace{1cm}

Thomas DeGrand,
Anna Hasenfratz,  and Tam\'as G.\ Kov\'acs\\
Department of Physics \\
 University of Colorado,
Boulder CO 80309-390

\vspace{1cm}

\mytoday \\ \vspace*{1cm}

\nopagebreak[4]

\begin{abstract}

In this paper we derive a simple parametrization of the cycling method developed
by us in our earlier work. The new method,
 called  renormalization group (RG) mapping, consists of a series of
carefully tuned APE-smearing steps. We study the relation between cycling and RG
mapping. We also investigate in detail how smooth instantons and
instanton-anti-instanton pairs behave under the RG mapping transformation. We use
the RG mapping technique to study the topological susceptibility and instanton size
distribution of SU(2) gauge theory. We find scaling in both quantities in a wide
range of coupling values. Our result for the topological susceptibility, 
$\chi^{1/4}=220(6)$ MeV, agrees with our earlier results.

\end{abstract}

\end{center}
\eject


\section{Introduction}

Instantons are an important part of the QCD vacuum. 
They are expected to play a major role in chiral symmetry breaking
and the low energy hadron spectrum \cite{Diakonov,Shuryak_long},
in addition to explaining the U(1) problem \cite{U1,WV}. 
 For the latter
the value and scaling properties of a global observable, the
topological susceptibility 
$$ \chi_{top} = {\langle Q^2 \rangle \over V} $$
is needed, while for describing chiral symmetry breaking the size and spatial
distribution of the instantons are  also important.

Lattice Monte Carlo simulation is the 
only available non-perturbative method to study the topology of 
the QCD vacuum from first principles. Since the instantons carry a
very small portion of the vacuum energy, on typical lattice configurations they are
hidden by the vacuum fluctuations.
This makes   their  study using lattice methods quite difficult. 
To overcome this problem, lattice studies either smooth the configurations by
some direct method while trying to preserve the underlying topological structure,
use  non-local observables that effectively and 
intelligently do the smoothing, or combine the two.
Examples for non-local observables are the 
the renormalization 
group inspired minimization method \cite{SPINMODEL,INSTANTON2}
and the methods that use fermionic observables and
relate the topological charge of a configuration to the index of the Dirac
operator through the Atiyah-Singer index theorem \cite{Narayanan,NS_PROC}.
Examples for direct smoothing are our  cycling \cite{SU2_DENS} procedure 
 or the different cooling techniques used by many
different groups \cite{COOL,MS,Forcrand}. 
 All of these methods have an intrinsic cutoff in the minimum instanton size
they can identify before it falls through the lattice. In 
addition,  nearby instanton-anti-instanton pairs might disappear from 
the lattice  during smoothing or
become otherwise invisible to the measuring algorithm. Nevertheless one expects
that as the continuum limit is approached and typical instanton sizes grow 
in lattice units, all these methods will give the same result. 

In several recent papers \cite{INSTANTON2,SU2_DENS}
we studied lattice topology using a method based on the
renormalization group equation. This method has the 
advantage of keeping instantons above a certain size 
of approximately one lattice spacing 
intact both in location and size, and  
preserving nearby pairs as well. However the method 
involves an ``inverse blocking'' step where
the original lattice is mapped through a minimization step into a fine lattice with
half the lattice spacing and twice the lattice size.
This makes  the method very slow and
memory intensive.
Using the RG method at fairly coarse lattice spacing, combined with a scale
invariant SU(2) fixed point (FP) action, we measured the topological susceptibility 
and the instanton size distribution.

We found $\chi^{1/4}=230(10)$ MeV and from the size distribution we 
concluded that the average  instanton in SU(2) has a radius of about 
0.2 fm. These values differ from the results obtained
by other groups. Both the improved cooling method \cite{Forcrand} and the 
``heating method'' of the Pisa group \cite{PISA_APE}
find a consistently smaller susceptibility, 
$\chi^{1/4}=198(8)$ MeV, which translates into
about a factor of 2 difference in the expectation value of $Q^2$. In addition,
the size distribution of Ref. \cite{Forcrand} peaks at about 
$\bar \rho = 0.43$ fm, more than twice our value.

It is essential to resolve the differences between the different methods. In this
paper we develop a procedure called renormalization group mapping (or RG mapping)
to approximate the cycling process of Ref.\ \cite{SU2_DENS}
 without actually constructing the fine
lattice. This way we can study larger lattices and perform more reliable scaling
tests. We have performed several tests,
 both on smooth instantons and Monte Carlo configurations,
 to understand and control the properties of the RG mapping method. 
 In order to compare our results directly with those
 of Refs. \cite{Forcrand,Delia}, 
we have also decided to use the Wilson action instead of our 
FP gauge action in this study.

All direct smoothing transformations distort the original lattice
configuration.  This makes the extraction of (continuum) short  to medium
distance physics,
like observations of topological objects, very delicate.
If any space-time symmetric smoothing transformation  
iis repeated
 enough times, all the vacuum structure in any finite
volume, including the simulation volume, will be washed away.
Thus, it does not make sense to extrapolate one's results to
the limit of a very large number   of smoothing steps.
The only measurements which are physically meaningful are measurements 
which are extrapolated back to  the original lattice, that is, back to
zero smoothing steps.  This requires careful monitoring of observables
over the whole history of smoothing transformations.

(The measurement of a long-distance observable, like a mass, is
more robust because one can design a smoothing algorithm which preserves
the spectrum of the transfer matrix.)

At the end the RG-mapping is a very simple and fast procedure for 
extracting the topological properties of a lattice configuration, but 
the price is that all observables must be monitored during the processing
and extrapolated back to their values at zero mapping steps.

The rest of the paper is organized as follows.
Section 2 describes the RG-mapping procedure and the method by which its
parameters are determined.
Section 3 shows tests of the RG-mapping method on lattice configurations.
Section 4 applies the RG-mapping method to measurements of the
topological susceptibility and the properties of instantons.
It also includes comparisons with previously published results by
us and others.
Finally, Section 5 is a summary.

\section {Cycling versus RG mapping}

The cycling process was discussed in Ref.\ \cite{SU2_DENS}
in detail. It consists of an inverse blocking transformation
followed by a blocking step mapping the original lattice to a lattice of the same
size and lattice spacing but greatly reduced vacuum fluctuations.
Our goal in this section is to develop an approximate mapping from the original
configuration to the cycled one without constructing the inverse blocked fine
lattice.

\subsection{Cycling}

The first step of a cycling transformation is an inverse blocking.
Inverse blocking identifies
the smoothest among the configurations that block back to the original configuration.
Technically,
 one has to solve the steepest descent RG fixed point equation
\bee
S^{FP}(V)=\min_{ \{U\} } \left( S^{FP}(U) +\kappa T(U,V)\right),  \label{STEEP}
\ee
where $\{U\}$ is the fine configuration with twice the size and half the lattice spacing
of the original $\{V\}$ configuration,
and $\kappa T(U,V)$ is the blocking kernel \cite{HN,PAPER1,PAPER2}.

While formally the inverse blocking is  non-local, for local FP actions the 
dependence of the fine
links on the original coarse links dies away exponentially
with their separation \cite{PAPER1} and the mapping
$\{V\} \to \{U(V)\}$ can be considered local. As a consequence, all long distance
properties of the coarse and fine configurations are identical, but the fine
configuration is  locally much smoother than the original one.

Ideally one would like to repeat the inverse blocking step many times until one
arrives to a configuration that still has all the long distance properties of the
original lattice but the vacuum fluctuations are reduced to a level that the vacuum
structure is clearly observable.

 Unfortunately this is not practical.
The only way to do repeated inverse blocking steps is to map the fine lattice at each
step to a smaller lattice  without changing the long distance properties or
introducing vacuum fluctuations.
Since the fine $\{U\}$ configuration has a very subtle coherent structure,
 a block transformation defined by the original blocking kernel $T$ of
Eqn.\ (\ref{STEEP}) would bring back most of the vacuum fluctuations. However any other
blocking will destroy the coherence and result in a smooth blocked configuration.
In Ref.\ \cite{SU2_DENS}  we used a shifted blocking. 
We did not change the blocking kernel but
shifted its origin to a different sublattice. The shifted transformation 
 has the unpleasant feature (for the present study) that a gauge
transformation of the original lattice does not correspond to a gauge
transformation on the cycled lattice. In other words, a gauge fixing on
 the original
lattice does not fix the gauge of the cycled lattice. This is not a problem if we
work with gauge invariant quantities like closed loops, but
in the RG mapping we present here,
 we are going to map the links of the original lattice to the
links of the cycled lattice, and therefore it is important that both lattices 
be in the same gauge.

It is easy to find a block transformation that 
preserves the gauge choice,
 but destroys the coherence of the fine lattice and gives a smooth cycled
configuration. In fact any block transformation that is based on the same
sublattice as the inverse blocking but has a different kernel than the inverse
blocking will do the job. For simplicity we chose a transformation where we first
replaced each link by an APE smeared link  \cite{APEBlock}
\beea
X_\mu(x) = (1-c)U_\mu(x) & +  & c/6 \sum_{\nu \ne \mu}
(U_\nu(x)U_\mu(x+\hat \nu)U_\mu(x+\hat \nu)^\dagger
\nonumber  \\
& + & U_\nu(x- \hat \nu)^\dagger
 U_\mu(x- \hat \nu)U_\mu(x - \hat \nu +\hat \mu) ),
\label{APE}
\eea
with  $X_\mu(x)$  projected back onto $SU(2)$.
Next on the APE transformed lattice we performed a blocking transformation
using the same kernel $T$ but with $\kappa$ equal to
infinity, as for the inverse blocking. 
The parameter $c$ is chosen to mimic the shifted blocking of Ref. \cite{SU2_DENS}.
 We found that this  modified
block transformation reduced vacuum fluctuations at the same rate as 
the shifted blocking when we chose the parameter c=0.75. 
It also gave the same topological structure as the shifted blocking 
on several configurations where we performed both procedures.

A cycling step now consists of the following: \\

\centerline{ 
$\{V\} -$ inverse blocking $\to \{U\} - $ APE smearing $\to \{X\} - $ 
 blocking $\to \{W\}$ ,}

and repeated cycling gives the series \\

\centerline{ 
$\{V\} \to \{W_1\} \to \{W_2\} \to ...\to  \{W_n\}$.}
The lattices $\{V\}$ and $\{W_n\}$ have the same size and lattice spacing but the
the vacuum fluctuations are greatly reduced.

\subsection{RG mapping} 

Our goal is to find a mapping from
$\{V\}$ to $\{W_n\}$ without doing the minimization of Eqn.\ (\ref{STEEP}).

Our new block transformation preserves the gauge. Therefore we require the mapping to
do the same. A gauge covariant way to construct the link variables
 $W_\mu(n)$ of
the cycled lattice is to fit them to a
 linear combination of different paths of the $\{V\}$
links connecting the points $n$ and $n+\mu$. We  used 14 different paths with length
up to nine links,
including the straight connection $V_\mu(n)$ and the length-three staples. From 
the 14 paths we formed up to 56 combinations and attempted to fit the links of
the $\{W\}$ lattice. For each link we fitted the 4 real numbers characterizing each
SU(2) matrix in the lattice
 independently, i.e. we did not constrain them to be an SU(2) element.
We  created about 20  $8^4$ lattices generated at lattice spacings
 $a\approx
0.11$ fm and
$a\approx 0.09$ fm for the fit. (We  used the FP action  of Ref.
\cite{SU2_DENS} at $\beta=1.6$ and
$\beta=1.7$.)  It turned out, however, that fits on the independent lattices
were consistent and the following fits have
 been obtained using one $a\approx 0.11$ fm and one
$a\approx 0.09$ fm lattices only.

First we start with the mapping $\{V\} \to \{W_1\}$. 
The best fit, using 56 operators, had  a standard deviation $\sigma^2\approx 0.014$ per
independent variable. If we first performed an APE-smearing $V\rightarrow X$
like Eqn. \ref{APE} (of course $U$ is replaced by $V$),
followed by a 56 parameter fit using the smeared $X$
 variables as links, the 
$\sigma^2$ was reduced to 0.008. The latter fit  
reproduced the value of the plaquette on the $\{W\}$
lattice better than 0.5\%  with a standard deviation per plaquette of about 0.02.
(The plaquette expectation value on the $\{W\}$ lattice is between 1.85-1.88.)
The topological charge (measured by the FP algebraic
operator \cite{SU2_DENS}) was found to be $6\times 10^{-5}$ on a charge-0 
configuration with standard deviation per site $1.4\times 10^{-3}$. 

Is such a complicated mapping really necessary? We  
tried the mapping $\{V\} \to \{W_1\}$ using several
consecutive APE-smearing steps like
Eqn. \ref{APE} (again, substitute $W$ for $X$ and $V$ for $U$).
The best fit had $\sigma^2\approx 0.01 $ .
The details of the APE
steps were not very important, as long as each step had $c<0.5$ and the sum of the
$c$ parameters added up to $\sum_i c_i \simeq 1.1$. This
simple fit reproduced the plaquette on the $\{W\}$
lattice up to 0.7\% and  gave
the topological charge as $1.2\times 10^{-4}$ for a zero charge
configuration. The corresponding standard deviations were 0.04 for the plaquette
and $1.8\times 10^{-3}$ for the topological charge. 
Since a sequence of simple APE transformations seems to work almost as well as a
complicated mapping, in the following we will concentrate on fits using a
sequence of APE-smearing transformations.

Next we map  $\{V\} \to \{W_2\}$. The procedure is similar as above.
The $\sigma^2$ of
this fit is around 0.01 using APE-smearing steps only.
Again, the goodness of the APE-smearing fit depended only on the sum of the
individual APE steps, $\sum_i c_i = 2.2$ giving the best $\sigma^2$.
For example, if we do 4 APE smearing steps with $c=0.45$ followed by a fifth step,
the best fit gives $c_5=0.4$ with $\sigma^2=0.01$. If we do 
5 APE steps with $c=0.4$ first, followed by
a fitted sixth step, we get $c_6=0.2$ with the same average deviation as before.

We constructed the mappings
$\{V\} \to \{W_n\}$ in a similar way. 
 In Table \ref{tab:APEpar} we
list the best parameters together with the appropriate $\sigma^2$ values. The
parameter  values at any given step give the sum of the required APE-smearing
steps $\sum_i c_i$. 
The exact
distribution of the $c_i$ parameters have very little effect on the fit quality
if each parameter $c_i\le0.5$. 
We found that a series of steps with $c=0.45$ followed by a 
last step with $c<0.45$ to complete the sum $\sum_i c_i$ is a good choice.

It might seem surprising that the quality of the fit actually improves as we fit
more times cycled configurations. 
All that means is that  the many times cycled configurations are
very smooth, and it is easier to fit a smooth configuration than the rough
quantum fluctuations of the few times cycled configurations. 

\begin{table*}[hbt]
\caption{Combined APE-smearing parameters to fit the cycled configurations.  }
\label{tab:APEpar}
\begin{tabular*}{\textwidth}{|@{}c@{\extracolsep{\fill}}|cccccc|}
\hline
mapping  to:\ \  &  $\{W_1\}$ & $\{W_2\}$ & $\{W_3\}$ & $\{W_4\}$ & $\{W_5\}$ &
$\{W_6\}$  \\
 \hline
$\sum c_i$ &  1.1  &  2.2      & 3.6 & 4.5 & 5.5 & 6.5    \\
\hline
$\sigma^2$   & 0.01 & 0.01 &  0.01 & 0.008 & 0.008 & 0.008  \\
\hline
\hline
mapping  to: \ \ &  $\{W_7\}$ & $\{W_8\}$ & $\{W_9\}$ & $\{W_{10}\}$ & $\{W_{11}\}$ &
$\{W_{12}\}$  \\
 \hline
$\sum c_i$ &  7.4 & 8.4  &  9.3      & 10.4 & 11.4 & 12.4    \\
\hline
$\sigma^2$   & 0.008 & 0.007 &  0.007 & 0.007 & 0.007 & 0.007  \\
\hline
\end{tabular*}
\end{table*}

\subsection{Comparing Cycling and RG mapping}

The ultimate test of the RG mapping is the comparison of the topological properties
of the cycled and RG mapped lattices. We  used the 20 $8^4$ lattices described
above for this test.
We  cycled the configurations up to 12 times and measured
the topological density using the algebraic operator defined in  \cite{SU2_DENS}.
We  also measured the total charge using the geometric operator. Next we RG
mapped the original configurations using consecutive APE steps each with parameter
$c\le 0.45$ as described in Table \ref{tab:APEpar}. 
We  analyzed the RG
mapped configurations the same way as the cycled ones.  

The instantons on these configurations have radii around $\rho=2a$, sometimes
even less. 
When instantons are  as small as these ones, their radius is quite uncertain. A
radius 1.5 instanton centered at a lattice site shows up quite differently from 
one centered in the middle of a hypercube.
We will analyze small smooth instantons in Section \ref{sec:smooth}.
Based on those results we note that instanton sizes 
around $\rho=2a$ are trustworthy only up to 10-20\%.

Here we consider in detail two sample configurations, a ``nice'' one and a
``complicated'' one. 
Configurations cycled less than 5-6 times are usually too rough to get a reliable
estimate of the instanton content.
Table \ref{tab:nice} lists the instanton content of the ``nice'' configuration after
6, 9 and 12 cycling and RG mapping steps.  The table also lists the topological
charge measured by the algebraic operator (the geometric operator gave
results which were consistent
with the algebraic operator)
 and the locations and radii of the instantons (I) and anti-instantons (A)
as well.
The instanton content of the cycled and fitted configurations are identical for
cycling steps 9 and 12 - the locations agree exactly and the radii agree within 10\%.
(Considering the difficulty in identifying small instantons, 10\% agreement in the
radii is actually better than expected.) On the 6 times RG mapped
configuration we found an extra instanton-anti-instanton pair. These two objects
are very close, about 2.5 lattice units or 50\% of their combined radii.
It is not obvious if they are the artifact of the instanton
finder on a rough configuration or if they are
 indeed present on the configuration 
but both cycling and RG mapping
annihilate them after a few steps. Since even in principle 
it is not possible to separate nearby pairs, we  are not worried about the
disappearance of this pair. In Section \ref{sec:smooth}
we will study the properties of pairs in  more detail.

\begin{table*}[hbt]
\caption{Instanton content of a ``nice'' $8^4$ configuration}
\label{tab:nice}
\begin{tabular*}{\textwidth}{@{}|c@{\extracolsep{\fill}}|c|l|c||c|c|c|}
\hline
  & \multicolumn{3}{c|} {Cycled} &  \multicolumn{3}{c|}{Fitted}  \\
\hline
cycling & $Q$ &$\rho$& location  & $Q$ & $\rho$ &location \\
\hline
6  & 0.002 & I\ \  2.1 & 0,7,2,1 & 0.002 & I\ \ 1.9 & 0,7,2,1 \\
   &       & A\ \ 2.2 & 7,1,1,6 &       & A\ \  2.0 & 7,1,1,6 \\ 
   &       &        &         &       & I\ \ 2.9 & 6,2,4,4 \\ 
   &       &        &         &       & A\ \  2.3 & 5,2,6,5 \\ 
\hline
9  & 0.001 & I\ \  2.3 & 0,7,2,1 & 0.001 & I\ \ 2.1 & 0,7,2,1 \\
   &       & A\ \  2.6 & 7,1,1,6 &       & A\ \ 2.3 & 7,1,1,6 \\ 
\hline
12  & 0.001 & I\ \  2.5 & 0,7,2,1 & 0.001 & I\ \  2.3 & 0,7,2,1 \\
   &       & A\ \ 2.8 & 7,1,1,6 &       & A\ \  2.6 & 7,1,1,6 \\ 
\hline
\end{tabular*}
\end{table*}

Table \ref{tab:complicated} lists the the instantons found on 
the ``complicated'' configuration.
That configuration has $Q=1$ on the cycled lattices. After 6 cycling steps
the instanton finder identifies 4 instantons and 3 anti-instantons. Three of the
instantons have radii less than 2. We can identify these instantons by their
location on the once cycled lattice where their radii are even smaller, 1.4, 1.5, 1.7,
respectively. These instantons grow slowly under the cycling transformation.
After 12 cycling steps we could identify only three of the four instantons
and only one of the three anti-instantons. The missing instanton and one of the
anti-instantons  at locations (4,7,0,0) and (3,7,2,7) 
formed a close pair with distance about 2.5 lattice units or 65\%
of their combined radii on the 6 times cycled configuration. 
It is quite possible that they have annihilated. 
The second missing anti-instanton is probably still present
as the topological charge on the lattice remained $Q=1$. 
It  most likely grew to a size comparable to the lattice itself ($\rho\approx 4$) and our
instanton finder could not locate it.
On the RG mapped configurations the topological charge was $Q=-2$. Comparing the
instanton content of the 6 times cycled/RG mapped configurations, we see that the
three small instantons did not survive the RG mapping transformation.
The 4 objects that survived 6 RG mapping steps appear to be stable, they are
present after 12 steps also.
Small instantons are obviously sensitive to local transformations and can get lost,
as this example shows.

\begin{table*}[hbt]
\caption{Instanton content of a  ``complicated'' $8^4$ configuration}
\label{tab:complicated}
\begin{tabular*}{\textwidth}{@{}|c@{\extracolsep{\fill}}|c|l|c||c|c|c|}
\hline
  & \multicolumn{3}{c|} {Cycled} &  \multicolumn{3}{c|}{Fitted}  \\
\hline
cycling & $Q$ &$\rho$& location  & $Q$ & $\rho$ &location \\
\hline
6  & 1.09 & I\ \  1.6 & 4,6,3,0 & -2.10 &     &  \\
   &      & I\ \  1.7 & 1,3,0,5 &      &     &  \\
   &      & I\ \  1.8 & 4,7,0,0 &      &     &  \\
   &      & I\ \  2.8 & 7,7,7,1 &      & I\ \  2.7 & 7,7,7,1  \\
   &      & A\ \  2.1 & 3,7,2,7 &      & A\ \  2.1 & 3,7,1,7  \\
   &      & A\ \  2.4 & 3,5,4,3 &      & A\ \  2.4 & 3,5,4,3 \\
   &      & A\ \  2.9 & 0,4,5,7 &      & A\ \  3.1 & 1,5,4,6 \\
\hline
12 & 1.04 & I\ \  1.9 & 4,6,3,0 & -2.10 &     &  \\
   &      & I\ \  2.0 & 1,3,0,5 &      &     &  \\
   &      & I\ \  2.9 & 7,7,0,2 &      & I\ \  3.0 & 7,7,0,2 \\
   &      & A\ \  2.6 & 4,5,3,3 &      & A\ \  2.5 & 4,5,3,3  \\
   &      &   &    &      & A\ \  2.3 & 2,7,1,7 \\
   &      &     &  &      & A\ \  3.7 & 2,4,6,6 \\
\hline
\end{tabular*}
\end{table*}

A similar analysis can be performed on all the other configurations. The qualitative
features are the same as on the sample configurations. If the instanton radius is
larger than about 1.5 lattice units on the original lattice,
the instanton is stable. It will survive the
cycling as well as the RG mapping steps. The location of the instanton is usually
unchanged (occasionally it is shifted
 by a lattice unit in one or more directions) 
and is identical on the RG
mapped and cycled lattices. The radii of RG mapped and cycled lattices agree
up to about 10\%, better for larger instantons.
The total topological charge is usually stable and identical on the cycled and RG
mapped lattices. When there is a difference, we can identify the small instanton
which was lost by one of the methods.
The disappearance of a small instanton sometimes creates a localized roughness on
the configuration. When that happens the algebraic topological charge operator
might give a value that is in between the old and new topological charge for a few
cycling/RG mapping steps before settling down at the new topological charge
value.  Apart from these ``transitional'' cases the algebraic charge is close to an
integer and it is also consistent with the geometric charge.

We have never observed the creation of an instanton between steps 6 and 12 as 
 suggested by Ref. \cite{NS_PROC}.

One feature that we  observe both on the cycled and RG mapped lattices is that
the instanton size changes slowly with
increasing number of smoothing steps. 
It usually increases. If we want to measure the size
distribution on the original lattice, we have to take this gradual change into
account.

\section{Properties of RG mapping}

As we have demonstrated in Section 1, a series of APE-smearing steps approximates
the cycling procedure closely. In this section we will study the behavior of
smooth instantons, pairs, and Monte Carlo generated instantons 
under RG mapping. We will also investigate how RG
mapping changes the short distance structure of the lattice.

As we pointed out in Section 1.2, the precise sequence of APE steps is not
important. In the following we perform 12-60 APE-smearing steps with parameter $c=0.45$
to emulate cycling steps 5-25. It is not important whether 
a certain number of APE
smearing steps, like 24, corresponds
 to 11 or 12 cycling steps or somewhere in between, as we are
monitoring the development of the configuration and will attempt to extrapolate it
back to the original lattice or zero APE-smearing.

\subsection{Properties of Smooth Instantons and Pairs under RG Mapping}
\label{sec:smooth}

We generated lattice configurations containing single instantons of 
various sizes by discretizing the continuum instanton solutions as
described in Ref. \cite{INSTANTON1}. Besides the instanton radius 
there is an additional parameter that might be important, especially
for small instantons. This is the location of the instanton center 
with respect to the lattice. We chose to study the two extremes,
i.e. instantons centered on a lattice site and centered in the 
center of a lattice cube. 

We performed
 up to 60 $c=0.45$ APE smearing steps on the single instanton configurations.
All instantons with $\rho/a>1.5$ survived 24 APE steps.
We found that if the instantons were centered on a lattice site, then
they could be safely identified even after 60 APE steps, provided their initial
radius was $\rho/a>1.25$. On the other hand for
instantons centered in the middle of a hypercube, this happened only
for $\rho/a>2.25$. This shows that the fate of small instantons can be 
very sensitive to their location with respect to the lattice.

\begin{figure}[htb!]
\begin{center}
\vskip 10mm  
\leavevmode
\epsfxsize=90mm
\epsfbox{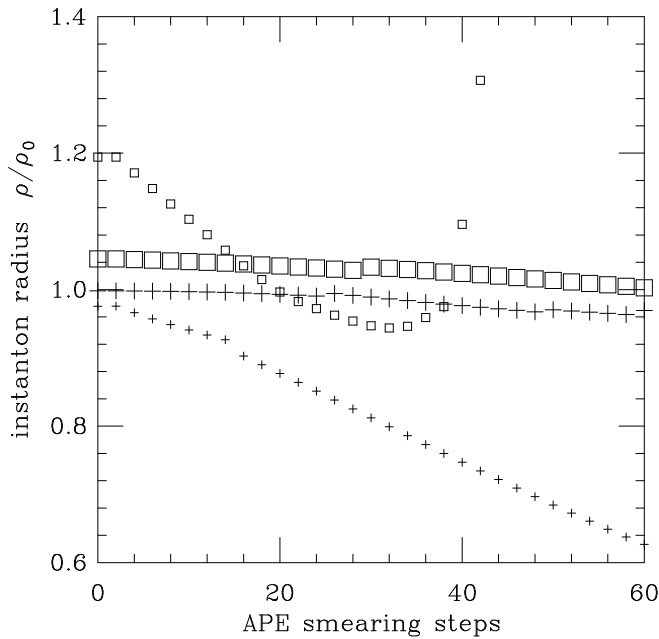}
\vskip 10mm
\end{center}
\caption{The ``APE history'' of a single instanton of radius $\rho=2.0$
(small symbols) and $\rho=4.0$ (large symbols) initially centered on a
lattice site (pluses) and in the middle of a hypercube (squares). The 
small instantons are on $8^4$ the big ones are on $16^4$ lattices.
The instanton radius is shown in units of the corresponding initial
continuum instanton radius i.e. 2 and 4 lattice spacings.}
\label{fig:single}
\end{figure}

As an illustration, in Fig.\ \ref{fig:single} we show the APE evolution 
of the size of single instantons
of radius 2 and 4 lattice spacings with the two different types of
locations for their centers. In order to avoid finite size effects
the radius 2 and 4 instantons were created on $8^4$ and $16^4$ lattices respectively. 
In all cases the instanton radius is shown in units of the corresponding
initial continuum instanton radius (i.e. 2 and 4). 
We can see that larger instantons tend to change less during the process.
For the $\rho/a=2$ instanton there is a 20\% difference in the
observed initial radius depending on the location of the center, while
for the $\rho/a=4$ instanton that uncertainty is reduced to below 5\%.
A linear extrapolation of the
size from 12-24 APE steps reproduces the 
size  on the original lattice within a few percent for all instantons.

The reason that we predict the size of instantons centered
at  lattice sites better that those centered
at the center of the hypercube is purely technical: our  instanton finding algorithm
compares every object to a smooth instanton centered at a lattice site.

We also observe that
that these hand crafted instantons always shrink under APE
smearing. This is expected since APE smearing is locally similar to
cooling with the Wilson action, which is expected to shrink the instantons
because smaller instantons have lower Wilson action. This is to be contrasted
with the observation that on real Monte Carlo generated configurations
only some of the instantons shrink, while most of them grow under APE smearing
(see Section \ref{sec:evolMC}).

We also did some tests on how oppositely charged pairs behave under APE smearing.
The pair configurations were produced by adding the logarithm of the links
of appropriately shifted single instanton 
configurations and then  re-exponentiating them.
As long as the separation of the pair was bigger than the sum of the radii
their sizes changed in essentially the same way as that of single instantons.
Finally we also looked at a few configurations with more instantons, their
locations and sizes taken from real many times cycled configurations. The
picture we found was the same again as with single instantons;
all the objects slowly shrank with a velocity depending on their size.

\subsection{Evolution of Monte Carlo Generated Instantons under RG Mapping}
\label{sec:evolMC}

Smooth instantons of size $\rho/a>1.5$ 
appear to be stable under RG mapping if no
more than 20-30 APE-smearing steps are done with $c=0.45$. The size of the
instantons decreases with RG mapping but a linear extrapolation to the original
lattice reproduces their size to 10-20\% accuracy for larger/smaller instantons.

The preliminary analysis of Section 2.3 shows that Monte Carlo generated
instantons of size $\rho/a>1.5$ are also stable under RG mapping but these instantons
usually grow under repeated APE-smearing.

\begin{figure}[htb!]
\begin{center}
\vskip 10mm
\leavevmode
\epsfxsize=90mm
\epsfbox{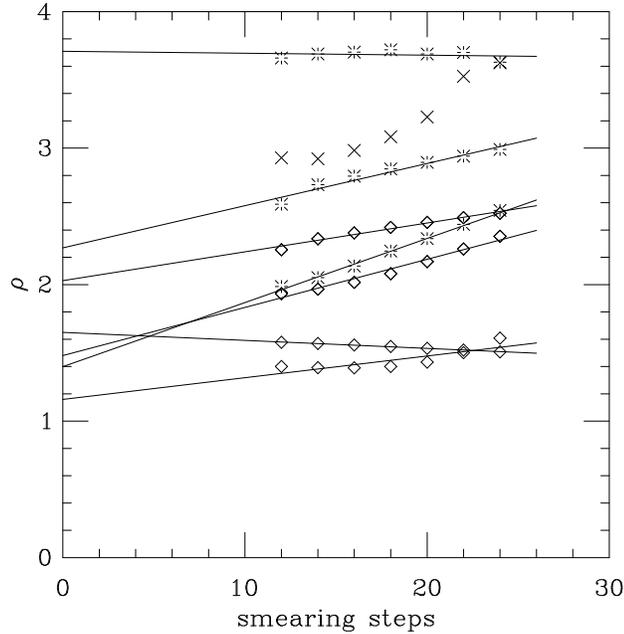}
\vskip 10mm
\end{center}
\caption{ Radius versus APE-smearing steps of instantons (diamonds) and
anti-instantons (bursts and crosses) on a $16^4$ $\beta=2.5$ configuration.}
\label{fig:rho_vs_APE}
\end{figure}

In this section we consider a $16^4$ configuration generated with
Wilson action $\beta=2.5$ (lattice spacing $a\approx 0.085$ fm) and study the
evolution of instantons between 12 and 24 $c=0.45$ APE-smearing steps.
 This configuration has a total charge $Q=2$ on all the studied configurations. 
After 12 smearing steps, the instanton finder identified 7 instantons and 6
anti-instantons. On the 18 times smeared configuration 
only 4 instantons and 4 anti-instantons
remain, and the same objects are observed even after  24 steps.
Fig. \ref{fig:rho_vs_APE} shows the size of the 8 ``stable'' objects as a function of the
APE-smearing steps. From the 4 instantons (diamonds) 3 increase in size
while one
decreases, but  all vary linearly with smearing steps.
The slope of the linear change for all of them is smaller than 0.035. Three of the
anti-instantons (bursts) behave similarly, though one has a slightly larger slope,
0.05. The fourth anti-instanton
(crosses) starts to grow rapidly after 18 smearing steps and will 
disappear after a few more steps. This object is likely to 
be a vacuum fluctuation, not an instanton.
Since the total charge is $Q=2$ on these configurations, either the instanton
identifier missed an instanton or one of the anti-instantons, probably the one with
the largest slope, is also only a vacuum fluctuation.

We can conclude that  the location of true 
topological objects is stable over many smearing steps
and their size changes slowly though it can increase or decrease as well. To identify
them on the lattice one has to track them over several smearing steps and monitor
their behavior.

\subsection{The Potential on RG Mapped Configurations}

\begin{figure}[htb!]
\begin{center}
\vskip 10mm  
\leavevmode
\epsfxsize=90mm
\epsfbox{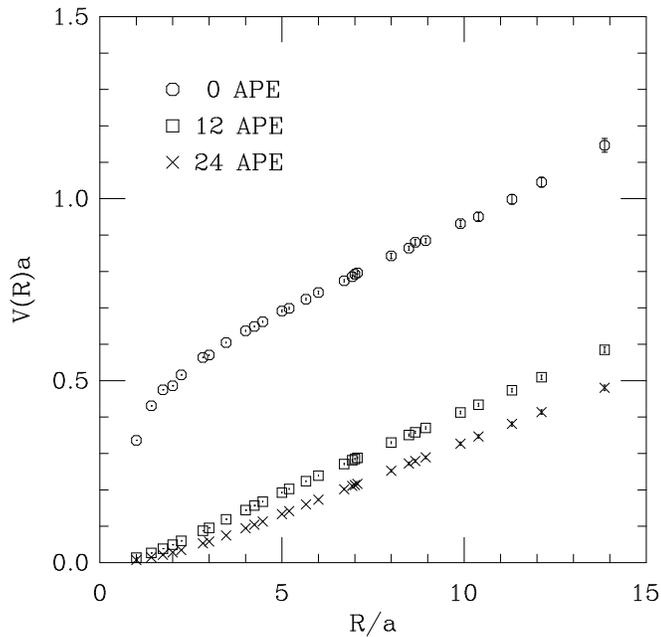}
\vskip 10mm
\end{center}
\caption{The heavy quark potential on a set of 120 $16^4$ configurations
at Wilson $\beta=2.5$ measured after 0 (octagons), 12 (squares) and 24
(crosses) $c=0.45$ APE smearing steps.}
\label{fig:pot_APE}
\end{figure}

In our previous work we saw that cycling did not change the long distance
features of the configurations and the string tension extracted
from the heavy quark potential was essentially unchanged through 
9 steps of cycling \cite{SU2_DENS}. Here we performed the same test 
for the RG-mapping. In Fig.\ \ref{fig:pot_APE} we show the heavy quark
potential extracted from timelike Wilson loops on the original 
configurations after 12 and 24 $c=0.45$ APE smearing steps. The measurement
was done on a set of 120 $16^4$ configurations generated at Wilson 
$\beta=2.5$. The qualitative features are the same as for cycling. 
As expected, the short distance part of the potential is changed
by APE smearing, and  the potential is shifted down, but the
string tension (the slope) is essentially unchanged.

\section{Results }

One of our goals in this paper is to compare results obtained with the cycling/RG
mapping method with results published using other algorithms
\cite{MS,Forcrand,Delia}.
Since most available data for topology was obtained using 
the Wilson  gauge action, we will
also use the  Wilson action in the following. 
It is known that the Wilson action has poorer
scaling in general and quite bad instanton scale invariance in particular. 
This is not a big problem if we compare results directly 
to others' calculations at identical coupling
values. As far
as scaling is concerned, we have to go to a large enough coupling where the
scaling violations are negligible.

\subsection{Parameters of the Simulation}

We have performed simulation at $\beta=2.4,2.5$ and $2.6$. The parameters of the
simulations are given in Table \ref{tab:parameters}. The configurations were
separated by 100 (at $\beta=2.4$), 150 (at $\beta=2.5$) and 200 (at $\beta=2.6$)
sweeps  and all our lattices are periodic. The lattice spacing in Table
\ref{tab:parameters} was calculated using the string tension results of
\cite{Fingberg_heller,Bali_su2} and the 
phenomenological formula of \cite{UKQCD_SU2}. Our statistics are
 about double that of Ref.\ \cite{Forcrand}.

\begin{table*}[hbt]
\caption{Parameters of the simulations}
\label{tab:parameters}
\begin{tabular*}{\textwidth}{@{\extracolsep{\fill}}|l|c|c|c|c|}
\cline{1-5}
$\beta$  & L  & number of & a [fm] & La [fm]  \\
         &   & configurations &  &          \\
\cline{1-5}
2.4 & 12 & 397 & 0.1210(1) & 1.45             \\
\cline{1-5}
2.4 & 20 & 28 & 0.1210(1) & 2.42             \\
\cline{1-5}
2.5 & 16 & 218 & 0.0850(5) & 1.36             \\
\cline{1-5}
2.6 & 20 & 370 & 0.061(2) & 1.22             \\
\cline{1-5}
\end{tabular*}
\end{table*}

\subsection{The Topological Susceptibility}

We have measured the topological charge using the FP algebraic operator
\cite{SU2_DENS} after 12, 18 and 24 $c=0.45$
APE-smearing steps. In addition, on the
$\beta=2.5$ configurations we  also measured the charge using the geometric topological
operator after 12 smearing steps.
Table \ref{tab:chi_sum} contains our results for $\langle Q^2 \rangle$ and $\chi^{1/4}$.
The quoted errors in $\chi^{1/4}$ are due to  the statistical 
error of $\langle Q^2 \rangle$. The error due to the 
uncertainty of the lattice spacing is negligible for $\beta=2.4$ and 2.5, but 
at $\beta=2.6$ it is about 3 per cent, significantly contributing to the
error of $\chi^{1/4}$. The second error at $\beta=2.6$ indicates the
 uncertainty due to the lattice spacing.
At $\beta=2.4$ we quote $\langle Q^2 \rangle$ only from the $12^4$ configurations. 
We have only 
28 configurations on the $20^4$ lattices. While the total volume of these
configurations is comparable to the
small lattice runs, it is not sufficient to get a statistically  meaningful 
value for the susceptibility.

\begin{table*}[hbt]
\caption{Results for the susceptibility }
\label{tab:chi_sum}
\begin{tabular*}{\textwidth}{@{}|l@{\extracolsep{\fill}}|c|c|c|c|}
\hline
$\beta$  & L  & smearing & $\langle Q^2 \rangle$ & $\chi^{1/4}$[MeV]   \\
\hline
2.4 &12 &                12 & 5.2(4) & 206(5)         \\
    &   &                18 & 4.8(4) & 204(5)         \\
    &   &                24 & 4.5(3) & 200(5)         \\
    &   & extrapolated to 0 & 5.9(4) & 211(5)         \\
\cline{2-5}
    & 8 & other& 0.86(13)& 197(7)        \\
\hline
2.5 & 16 &          12(geom) & 5.4(5) & 221(5)       \\
    &    &           12(alg) & 5.4(5) & 221(5)       \\
    &    &                18 & 5.2(5) & 218(6)       \\
    &    &                24 & 5.1(5) & 217(6)       \\
    &    & extrapolated to 0 & 5.5(5) & 222(6)       \\
\hline
2.5115    & 12 & other    &1.5(2) & 221(7)               \\
\hline
2.6 & 20 &                12 & 3.3(3) & 218(8)(7)        \\
    &    &                18 & 3.3(3) & 218(10)(7)       \\
    &    &                24 & 3.2(2) & 216(10)(7)       \\
    &    & extrapolated to 0 & 3.4(3) & 220(10)(7)       \\
\cline{2-5}
    & 16 & other & 1.26(15) & 214(7)(7)   \\
\hline
\end{tabular*}
\end{table*}
 
We observe, as expected, a 
small systematic decrease in the susceptibility as we increase the
number of smearing steps.
At large $\beta$ the change is small and statistically insignificant. 
Only at $\beta=2.4$, where the configurations are the roughest, 
do the 12 and 24 smearing steps results differ by about  a standard
deviation. 

\begin{figure}[h!tb]
\begin{center}
\vskip 10mm
\leavevmode
\epsfxsize=90mm
\epsfbox{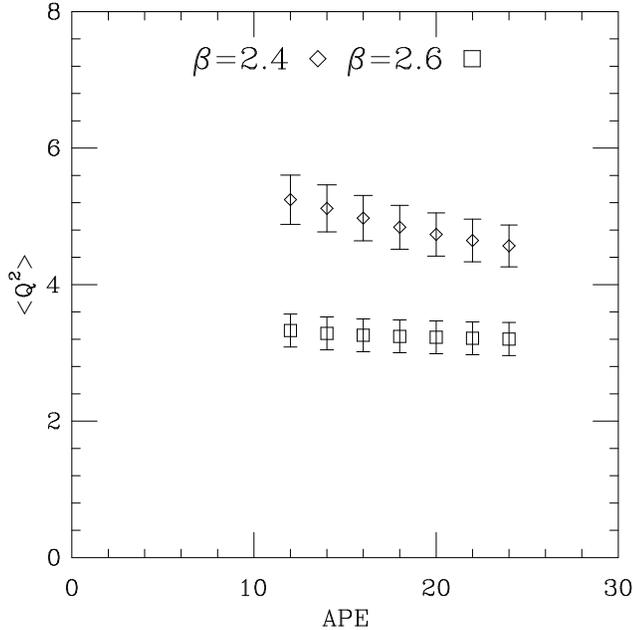}
\vskip 10mm
\end{center}
\caption{The susceptibility $\langle Q^2\rangle$ vs number of
$c=0.45$ APE  steps.  Symbols are diamonds for $\beta=2.4$ and squares
 for $\beta=2.6$.}
\label{fig:chi_vs_sweep}
\end{figure}

Fig.\ \ref{fig:chi_vs_sweep} shows the variation in 
$\langle Q^2\rangle$ with smoothing steps at $\beta=2.4$ and $2.6$.
The change occurs because instantons are lost under smoothing.
Since the susceptibilities measured on the same set
of configurations after different numbers of smearing steps are 
correlated, in spite of the statistical errors it is still meaningful 
to extrapolate them to 0 smearing step. These extrapolated values are
also quoted in Table \ref{tab:chi_sum}.
With the extrapolation we take into account the contribution of instantons that are lost
during the smoothing steps, either because they were too small, and 
smoothing washed them away,
or because they were too large, and during repeated smoothing they grew out of the lattice.
 The extrapolation cannot account for instantons that are too small to be present
even on the original, unsmoothed lattice.
The susceptibility increases by 10\% from $\beta=2.4$ to $\beta=2.5$
but stabilizes after that at the value $\chi^{1/4}=220(6)$ MeV.
We interpret the change between  the extrapolated $\beta=2.4$ result
and the larger $\beta$ results as due to the absence of
small  instantons  at $\beta=2.4$ because of the larger lattice spacing. This
interpretation will be supported by the instanton size distribution result
discussed in Section \ref{sec:size_dist}. 
The result for the susceptibility is consistent with our earlier results
\cite{INSTANTON2},\cite{SU2_DENS} where we found $\chi^{1/4}=230(10)$
MeV using the cycling method with the FP action on much rougher lattices. 

Table \ref{tab:chi_sum} contains another set of results for $\langle Q^2 \rangle$ and
$\chi^{1/4}$ marked ``other''. Those were obtained using an earlier, 
unpublished version of the RG
mapping algorithm. That mapping used about 50 paths to map the original
configuration to the cycled one obtained with the shifted blocking procedure. We
have not tested that algorithm in such  detail as the present one, but
preliminary results indicate that it has about the same minimum instanton size as
the present algorithm. The agreement between the results supports this expectation
and also indicates that finite size effects are indeed small.

We can compare the data for $\langle Q^2 \rangle$ with the result published in Ref.\
\cite{Forcrand}. They quote the susceptibility after 20, 50, 150 and 300 improved
cooling steps. The general trend is that the susceptibility slowly decreases 
during cooling but the change slows down with more cooling. For example
at $\beta=2.4$ the ``cooling history'' of $\langle Q^2 \rangle$ is 
4.5(3), 4.4(3), 4.2(3), 4.0(3) for 20, 50, 150 and 300 cooling steps.
The change in the susceptibility is comparable or smaller than its
statistical error and the authors of \cite{Forcrand} consider that $\langle Q^2 \rangle$
stabilizes after $O(100)$ cooling steps.
At $\beta=2.4$ on the same lattice size as Ref. \cite{Forcrand} we found
$\langle Q^2 \rangle=4.5(3)$ after 24 APE steps. This result is 
somewhat larger than, 
but  consistent with  \cite{Forcrand}. However our zero smearing step 
extrapolated number,
$\langle Q^2 \rangle=5.9(4)$ is several standard deviations different from the result of
\cite{Forcrand} even if we consider their $\langle Q^2 \rangle$ after 20 cooling sweeps. 
At $\beta=2.5$ only a small data set on $12^4$ lattices was used in \cite{Forcrand}. The
published result for the susceptibility is $\chi^{1/4}=217(7)$ MeV indicating
$\langle Q^2 \rangle=1.45(20)$. Rescaling with the volume so 
it corresponds to our lattice size predicts
$\langle Q^2 \rangle|_{16^4}=4.6(6)$, again lower 
than, but consistent with, our 24 APE steps
result but two standard deviation different than our extrapolated result.
At $\beta=2.6$ in Ref.\ \cite{Forcrand} $24^4$ lattices with
twisted boundary condition were used. The published result for the charge is
$\langle Q^2 \rangle=4.2(7)$ (after 150 cooling steps).
Rescaling with the volume factor translates to $\langle Q^2 \rangle|_{20^4}=2.0(3)$ 
on a $20^4$ lattice. This number is 50\% lower and 
3 standard deviations away from ours.  Finite size effects cannot 
explain the difference as they would push our numbers even higher. We do not
 understand the reason of this discrepancy, especially not why it shows up
only at the largest coupling value.

In the remainder of this section we will discuss the relation between the
algebraic and geometric definition of the topological charge and the possibility
of a non-trivial renormalization group factor in the definition of the topological
susceptibility \cite{PISA_APE}.

\begin{figure}[h!tb]
\begin{center}
\vskip 10mm
\leavevmode
\epsfxsize=90mm
\epsfbox{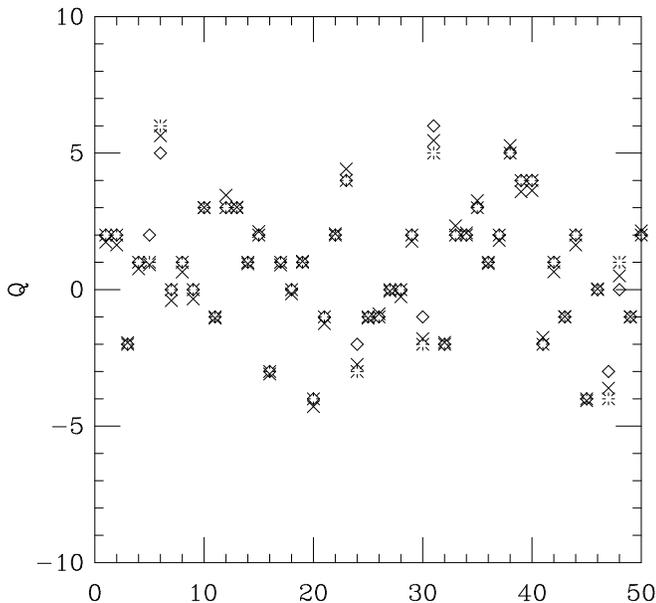}
\vskip 10mm
\end{center}
\caption{The geometric (diamonds) and algebraic (crosses) charge on a sequence of 
$\beta=2.5$ 12 times smeared configurations. Since the algebraic charge is not
integer, for clarity we plot the nearest integer value (bursts) also.}
\label{fig:geom_and_alg_int}
\end{figure}

First, we
 compare the geometric and algebraic charges on the 12 times smeared
$\beta=2.5$ lattices. $\langle Q^2 \rangle$ 
agrees with the two methods, but we can compare the
charges measured on the individual configurations as well. Since the algebraic
charge does not give an integer value, we compare the geometric charge to
the nearest integer of the algebraic charge for the first 50 of the $\beta=2.5$
configurations (Figure \ref{fig:geom_and_alg_int}). Here the diamonds correspond
to the geometric charge and the bursts to the integer algebraic charge. 
The two agrees most of the
time, occasionally differing by one. 
The crosses in Figure \ref{fig:geom_and_alg_int} correspond to the original,
non-integer value. They are usually close to an integer, and when the geometric
and the integer algebraic charges  are different, the original value is in
 between the two.

Just how close is the algebraic charge to an integer?
To quantify that, we use the following three quantities
\beea
\langle (\Delta Q)^2 \rangle = \langle (Q^{int}-Q)^2 \rangle, \nonumber \\
Z^2= \langle Q/Q^{int} \rangle \;\; \mbox{if} \;\;\; {Q^{int} \ne 0}, \\
M= \langle Q^2 \rangle \;\;\; \mbox{if} \;\;\; {Q^{int}=0}, \nonumber
\eea
where $Q^{int}$ is the nearest integer to the algebraic charge  $Q$.
The last two quantities correspond to the renormalization factors of Ref.\
\cite{PISA_APE,Delia} if we assume $Q^{int}$ is the true topological charge.
It is expected that a scale invariant charge operator which gives
close to integer values on smooth instantons, as ours does, has $Z=1,M=0$ on
highly smoothed configurations.
At $\beta=2.5$  after 24 smearing steps we find
\bee \langle (\Delta Q)^2 \rangle =0.03(1), \ \ Z^2=1.01(1), \ \ M=0.02(1),
\ee 
and at $\beta=2.6$
\bee \langle (\Delta Q)^2 \rangle =0.02(1), \ \ Z^2=1.02(1), \ \ M=0.01(1),
\ee 
consistent with this expectation.
Since the measured susceptibility is consistent after 12,18 and 24 smearing steps,
we expect similar results  for $Z^2$ and $M$ after 12 steps as well. 
Indeed, after 12 smearing steps the corresponding values  at $\beta=2.5$ are
\bee \langle (\Delta Q)^2 \rangle =0.07(2), \ \ Z^2=0.98(1), \ \ M=0.07(2) ,
\ee
and at $\beta=2.6$
\bee \langle (\Delta Q)^2 \rangle =0.04(2), \ \ Z^2=1.01(1), \ \ M=0.04(1) ,
\ee
indicating that if there is any renormalization necessary of the topological
susceptibility, it is very small.
The above argument is based on the assumption that $Q^{int}$ is the true charge of
the configuration. We cannot disprove the possibility that even though the
algebraic topological charge is close to an integer value 
($ \langle (\Delta Q)^2 \rangle =0.03-0.07$), the true 
topological charge differs from it by some overall multiplicative 
renormalization factor, but we find this possibility unlikely.

\subsection{The Instanton Size Distribution}
\label{sec:size_dist}

We have identified individual instantons after every 2 APE-smearing steps between 12 and 24
steps. Since these configurations are still rough, many of the objects identified as
instantons are in fact vacuum fluctuations and disappear after more smoothing steps. This is
best illustrated if we calculate the topological susceptibility obtained using $I-A$, the
difference between the number of instantons and 
anti-instantons on  a given configurations as
the charge operator. For example at $\beta=2.5$ we find 
$\langle (I-A)^2 \rangle =11.1$ after 12, 8.8 after
18, and 8.3 after 24 smearing steps - 1.5 to 2 times larger than the direct measurement of
$\langle Q^2 \rangle$ on the same lattices.

\begin{figure}[h!tb]
\begin{center}
\vskip 10mm  
\leavevmode
\epsfxsize=90mm
\epsfbox{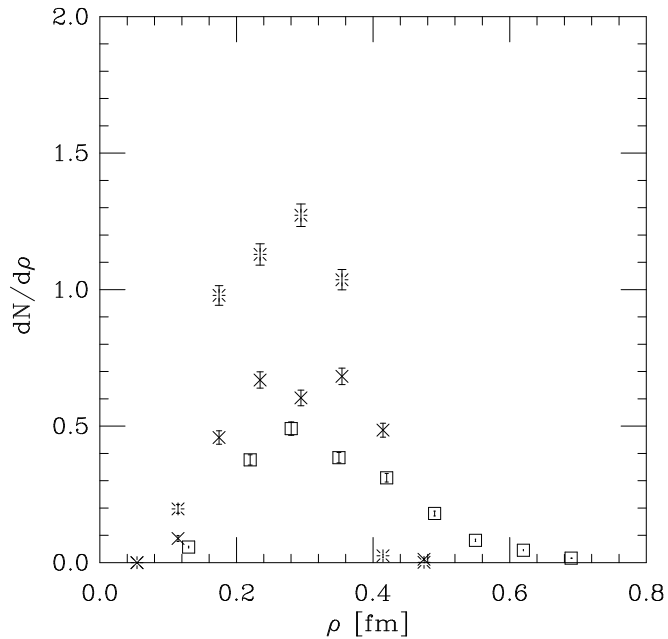}
\vskip 10mm
\end{center}
\caption{The size distribution on 12 (bursts) and 24 (crosses) times APE smeared lattices at
$\beta=2.5$. The square symbols are the results of Ref.\ \cite{MS} rescaled appropriately.}
\label{fig:d_rho_direct}
\end{figure}

An option is to smooth the lattice further. However we would like to keep the smoothing
steps small to minimize the distortion of the lattice, like losing instantons and
changing the size of instantons. Figure \ref{fig:d_rho_direct} shows the observed instanton 
size distribution on the 12 and 24
times smeared lattices at $\beta=2.5$. For comparison we also plot the result of Ref.
\cite{MS} which corresponds, in our normalization,
 to about 100 times smeared lattices. It is
obvious from the figure that the total number of identified objects decreases
 as we increase
the smoothing. 
The density of identified objects is 4.6 per  fm$^4$ after 12
smearing steps, 3.0 fm$^4$ after 24 
smearing steps, and about 2 per fm$^4$ in Ref. \cite{MS}. These density values are
considerably larger than the expected value of about 1 per fm$^4$.
The maximum of all 3 distributions is around $\bar \rho \approx 0.3$ fm, but
that does not mean that on the original lattice  $\bar \rho \approx 
0.3$ fm since instantons usually grow under smearing. This 
growth can also  be observed from the increasing tail of the
distribution, especially after the many blocking steps of \cite{MS}.

What are these objects that disappear after repeated smearing steps but contaminate the
result if we smear only 20-30 times? A few of them are genuine instantons, but most of them
are vacuum fluctuations that the instanton finder erroneously identifies as topological
objects. Some of these vacuum fluctuations could be nearby instanton - anti-instanton pairs.
The topological susceptibility from $\langle (I-A)^2 \rangle$ is quite different form
the reliable, direct measurements, indicating that most of the extra objects are only vacuum
fluctuations.
Our method to distinguish true topological objects from vacuum fluctuations is to track
individual objects during smearing. If an object is present over many smearing steps with
slowly changing radius, we identify it as a topological object.

At smaller coupling values $\beta=2.4-2.5$ we found that 12 to 18 
$c=0.45$ smearing steps were
sufficient, but at $\beta=2.6$ we needed 16 to 
24 steps to smooth the configuration so larger
instantons emerged from the vacuum fluctuations. The necessary smearing steps would increase
further if we were to consider even larger couplings but as large instantons are affected
less by the smoothing, that would not influence the result.

In the following analysis we identified instantons on configurations that had been APE
smeared 16, 18, 20, 22 and 24 times with $c=0.45$,
and extrapolated their size linearly to zero smearing
level. We kept only instantons where the slope of the linear extrapolation was less that
0.035. This cut was chosen such that the final distribution at each $\beta$ value
reproduced the topological susceptibility correctly, 
$\langle(I-A)^2\rangle=\langle Q^2 \rangle$.
With this cut the instanton density is 1.2 per fm$^4$ at $\beta=2.4$ and 1.5 per fm$^4$  at
$\beta=2.5$ and 2.6. Again, we see that at the smallest coupling value  
some instantons are missing.
These densities are 15-20\% lower than the one we quoted in \cite{SU2_DENS}. A
reason for the discrepancy is that in \cite{SU2_DENS} we used a  50\% larger 
cut in the
slope of the linear extrapolation. (Our statistics was not sufficient to adjust 
the cut to give the correct topological
susceptibility.)
We want to emphasize that our result for the instanton density, 1.5 per fm$^4$ is not a
direct measurement. It is the result of combining 5 different APE smeared lattices and
identifying those objects that are present and stable on all of them.

It is interesting to compare these densities with the results of Ref. \cite{Forcrand}. Table
3.3.2 of \cite{Forcrand} contains raw data for the instanton distribution from which one can
calculate the instanton density. At $\beta=2.4$ after 20 cooling sweeps the density is 2.5
per fm$^4$, and after 300 sweeps it is 0.5 per fm$^4$. A density
 of 2.5 per fm$^4$ is quite a
bit larger than the phenomenologically expected value and probably contains
 a lot of vacuum
fluctuations or instanton-anti-instanton pairs which are too close to
 distinguish them from
vacuum fluctuations. On the other hand the density 0.5 per
 fm$^4$ is 2-3 times smaller than
expected, indicating that cooling destroyed not only vacuum 
fluctuations but a lot of topological objects as well. It is not obvious to us
why these two very different sets of configurations have the same 
instanton size distribution as it was found in Ref. \cite{Forcrand}.
The densities at $\beta=2.6$ show a similar trend. After 50 sweeps
 the density is large, 3.7
per fm$^4$, and it drops to 0.9 per fm$^4$ after 300 sweeps. 

\begin{figure}[h!tb]
\begin{center}
\vskip 10mm
\leavevmode
\epsfxsize=90mm
\epsfbox{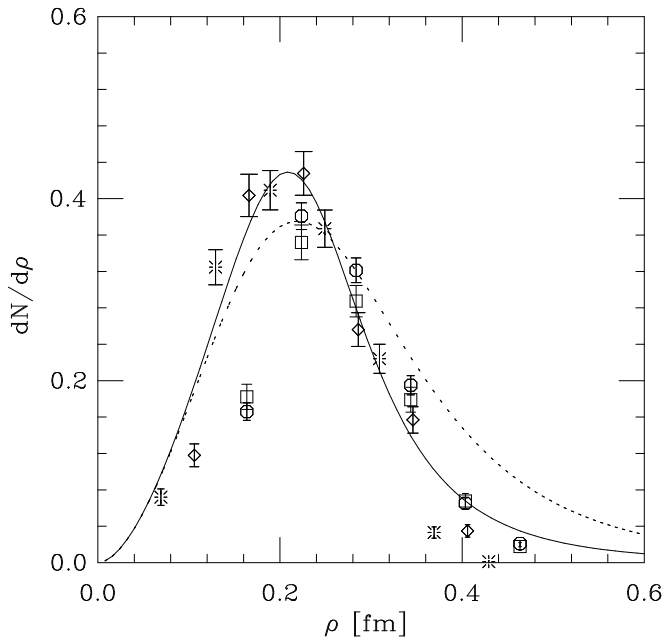}
\vskip 10mm
\end{center}
\caption{The size distribution of instantons. Octagons correspond to $\beta=2.4$ $12^4$,
squares  to $\beta=2.4$ $20^4$, diamonds
to $\beta=2.5$ and bursts to $\beta=2.6$. The first bin of each distribution is
contaminated by the cut-off. 
The solid curve is a two parameter fit to the data points according to the formula 
in Ref.\ \cite{Shur.distrib} The dashed curve is a similar fit from Ref.\ 
\cite{Shur.distrib} which describes the instanton liquid model quite closely. }
\label{fig:d_rho_wil}
\end{figure}

Our final result for the instanton size 
distribution is shown in Figure \ref{fig:d_rho_wil}, where
we overlay the data obtained at $\beta=2.4$ (octagons on $12^4$, squares on $20^4$ lattices), 
$\beta=2.5$ (diamonds), and
$\beta=2.6$ (bursts). The bin size is 0.06 fm and the distribution $d\rho / dN$ 
is measured in fm$^{-4}$. Since the smoothing method cannot identify instantons with
$\rho<1.5a$, we chose the bins such that the second bin for each distribution starts at
$\rho=1.6a$. That means that we expect the second bin of each distribution to be universal,
which appears to be true  where we can check it, i.e. 
for the $\beta=2.4$ and 2.5 distributions. The first bins on the
other hand contain only some of the small instantons and their value is not expected to be
universal.
 
The four distributions form a universal curve 
indicating scaling. The $\beta=2.4$ curves cover
only the $\rho>0.2$ fm region, 
and small instantons are obviously missing. The agreement between the $12^4$ and $20^4$
configurations at $\beta=2.4$ indicate that a linear size of about 1.4 fm is sufficient to
observe all the topological objects. The $\beta=2.5$ and
2.6 distributions have most of the physically relevant instantons supporting the scaling
behavior observed for the topological susceptibility.
It is possible, though statistically not significant, that we see some 
finite size effects for $\rho>0.36$ fm at the $\beta=2.6$
distribution. It is plausible that objects which
 are larger in diameter than about 0.8 fm will
grow and after 16-24 APE-smearing steps become too big to be observed
on a lattice of size 1.22 fm.

Our instanton size distribution is very different from the result of Ref. \cite{Forcrand}.
Their  distribution peaks at $\rho=0.43$ fm and they found hardly any 
objects with $\rho<0.2$ fm.
However, the instanton liquid model predicts a very similar picture to ours. 
In the interacting instanton liquid model Shuryak predicted an instanton size 
distribution that peaked
around $\rho=0.2$ fm \cite{Shur.distrib}. In that paper the instanton distribution of
\cite{MS} was fitted using the two loop perturbative instanton distribution formula with a
``regularized'' log
\beea
S_I={8\pi^2 \over g^2(\rho)} = b_0L + b_1 \log L \\
L={1 \over p}\log[(\rho\Lambda_{inst})^{-p}+C^{p}] 
\eea
where $b_0$ and $b_1$ are the first two coefficients of the 
perturbative $\beta$ function and $p$
and $C$ are arbitrary parameters.
For SU(2) 
$\Lambda_{inst}=0.66$ fm and the best fit for the data 
of Ref.  \cite{MS} gives $p=3.5$, $C=4.8$. 
The best fit for our data requires different values, $p=6.4$, $C=5.95$. The solid
curve in Figure \ref{fig:d_rho_wil} corresponds to the latter fit while the dashed curve is
the fit given in Ref. \cite{Shur.distrib} that describes the instanton liquid model quite
closely.  The
difference between the two fits is significant at large $\rho$ values only. Our
distribution and the corresponding fit cuts off much faster at large instanton sizes than
the instanton liquid model prediction.  A natural explanation would be that finite size
effects in our simulations cause the Monte Carlo distribution to drop too fast at large
$\rho$. However we compared distributions at linear lattice
sizes 1.44 fm and 2.42 fm and observed no difference, 
so this explanation does not appear to be true.
We do not know if changing the parameters of the interacting instanton 
liquid model slightly would change
the predictions of that model improving the agreement with the Monte Carlo data.

In Ref.\ \cite{SU2_DENS} we published a similar size distribution curve obtained by the
cycling method on much rougher lattices using the FP action. That distribution agrees with
the present one for $\rho<0.35$ fm but is slightly larger for $\rho>0.35$ fm. The reason for
this difference is probably  the same  that caused the slightly larger overall density.

\section{Summary of the RG Mapping Method}
 
Our RG mapping method is a series of APE-smearing transformations. With
smearing parameter $c=0.45$ we found that the topological charge can be reliably
measured after 12 to 16 smearing steps  (slightly more at smaller $\beta$ values).
Individual instantons can be identified also after about 12 steps but to reliably
distinguish them from vacuum fluctuations it is necessary to monitor them over
several smearing steps. Since instanton sizes change during RG mapping,
tracking them over several smearing steps is also necessary so we can extrapolate
their size to the original lattice.
The smearing steps destroy the short distance properties of the lattice. The 12 to 24
steps that we are doing destroys structure on 
the lattice up to a few lattice spacings. As the
lattice spacing decreases, this lattice short distance distortion becomes
negligible.

APE smearing as a smoothing method for topology
 is not new. The Pisa group \cite{PISA_APE} has
been using it for several years now. They usually do 2 APE-smearing steps with
$c\approx 0.9$.  This
  corresponds to about 4 of our smearing steps or two RG mapping
steps. These configurations are not smooth enough for us  to measure the
topological charge directly. The Pisa group corrects for the vacuum
fluctuations by a multiplicative and an additive
renormalization factor. These renormalization factors also, in principle,
partially correct
  for the fact that the naive twisted plaquette $F\tilde F$ operator
used in most of their studies, gives
the topological charge of small instantons quite poorly \cite{INSTANTON1}.
 
The ``calibrated cooling'' method of Michael and Spencer \cite{MS}
  is
very similar to APE-smearing, except that the lattice is updated continuously
during the transformation. They chose a smearing parameter $c=0.75$ and performed
60-80 steps. That corresponds to over 100 steps in our notation. That many steps
destroys many more instantons and most importantly,
the size of the instantons can grow considerably.
$O(100)$ smearing steps can also influence the short distance properties of the
lattice considerably.
 
Our method differs from the above two in the following:
\begin{itemize}
\item{It was motivated by and tested against the RG cycling
transformation.} 
\item{We do enough smearing steps to be able to use an algebraic charge
operator that gives close to integer values. }
\item{We monitor that the long distance properties of the lattice are
unchanged. The smearing steps destroy correlations at only a few (1-2)  lattice
spacings.}
\item{We monitor individual instantons over several smearing steps to  
distinguish them from vacuum fluctuations. We also monitor their size and
extrapolate it back to the original lattice.} 
\end{itemize}

We used Wilson action at several coupling values in our numerical 
simulation. We found that the topological susceptibility showed scaling 
for $\beta\ge 2.5$, and $\chi^{1/4}=220(6)$ MeV. The instanton
density (instantons and anti-instantons combined) is 1.5 fm$^{-4}$, and 
the instanton size distribution peaks at $\bar\rho=0.2$ fm as shown 
in figure \ref{fig:d_rho_wil}. A similar technique for $SU(3)$ gauge theory 
is under development. This method of removing short distance fluctuations is 
very fast and easy to implement, but its price is that all observables must be 
carefully monitored over the history of processing steps and, 
if necessary, extrapolated back to zero  steps.

\section*{Acknowledgements}
We would like to thank  the Colorado High Energy experimental
groups for allowing us to use their work stations.
This work was supported by the U.S. Department of Energy
 grant DE--FG03--95ER--40894.

\newcommand{\PL}[3]{{Phys. Lett.} {\bf #1} {(19#2)} #3}
\newcommand{\PR}[3]{{Phys. Rev.} {\bf #1} {(19#2)}  #3}
\newcommand{\NP}[3]{{Nucl. Phys.} {\bf #1} {(19#2)} #3}
\newcommand{\PRL}[3]{{Phys. Rev. Lett.} {\bf #1} {(19#2)} #3}
\newcommand{\PREPC}[3]{{Phys. Rep.} {\bf #1} {(19#2)}  #3}
\newcommand{\ZPHYS}[3]{{Z. Phys.} {\bf #1} {(19#2)} #3}
\newcommand{\ANN}[3]{{Ann. Phys. (N.Y.)} {\bf #1} {(19#2)} #3}
\newcommand{\HELV}[3]{{Helv. Phys. Acta} {\bf #1} {(19#2)} #3}
\newcommand{\NC}[3]{{Nuovo Cim.} {\bf #1} {(19#2)} #3}
\newcommand{\CMP}[3]{{Comm. Math. Phys.} {\bf #1} {(19#2)} #3}
\newcommand{\REVMP}[3]{{Rev. Mod. Phys.} {\bf #1} {(19#2)} #3}
\newcommand{\ADD}[3]{{\hspace{.1truecm}}{\bf #1} {(19#2)} #3}
\newcommand{\PA}[3] {{Physica} {\bf #1} {(19#2)} #3}
\newcommand{\JE}[3] {{JETP} {\bf #1} {(19#2)} #3}
\newcommand{\FS}[3] {{Nucl. Phys.} {\bf #1}{[FS#2]} {(19#2)} #3}


\end{document}